# Branch prediction related Optimizations for Multithreaded Processors

Murthy Durbhakula

Indian Institute of Technology Hyderabad, India

cs15resch11013@iith.ac.in, murthy.durbhakula@gmail.com

**Abstract** Major chip manufacturers have all introduced Multithreaded processors. These processors are used for running a variety of workloads. Efficient resource utilization is an important design aspect in such processors. Depending on the workload, mis-speculated execution can severely impact resource utilization and power utilization. In general, compared to a uniprocessor, a multithreaded processor may have better tolerance towards mis-speculation. However there can still be phases where even a multi-threaded processor performance may get impacted by branch induced mis-speculation. In this paper I propose monitoring the branch predictor behavior of various hardware threads running on the multi-threaded processor and use that information as a feedback to the thread arbiter/picker which schedules the next thread to fetch instructions from. If I find that a particular thread is going through a phase where it is consistently mis-predicting its branches and its average branch misprediction stall is above a specific threshold then I temporarily reduce the priority for picking that thread. I do a qualitative comparison of various solutions to the problem of resource inefficiency caused due to mis-speculated branches in multithreaded processors. This work can be extended by doing a quantitative evaluation.

## 1 Introduction

Major chip manufacturers have all introduced multithreaded processors. Such chips are used for running various multi-threaded and multi-programmed workloads. Efficient resource utilization is an important design aspect on such processors. There are various stalls that can impact resource efficiency such as DRAM access stalls, cache-to-cache transfer stalls, and branch mis-speculation stalls. Of these branch mis-speculation stalls are most difficult to predict as branch predictors have become quite sophisticated [1]. In this paper I propose a mechanism to monitor phase based behavior of such branch stalls and provide feedback to the arbiter which picks threads to be scheduled on the multi-threaded processor. The arbiter can use that information to lower the priority of the thread impacted by branch stalls thereby increasing both resource utilization and power savings. The rest of the paper is organized as follows: Section 2 presents the mechanism to manage branch predictor stalls. Section 3 briefly describes the qualitative methodology I used in evaluation. Section 4 presents results. Section 5 describes related work and Section 6 presents conclusions.

## 2 Managing branch predictor stalls

For every thread we maintain a "Branch Misprediction Counter" which is a running count of branch mispredictions in a given time window T and "Branch Mis-prediction Stall Cycles" counter which is a running count of branch misprediction stall cycles in a given time window T. We define a metric called AverageBranchMispredictStall = (Branch Misprediction Stall Cycles Counter)/(BranchMispredictionCounter). We calculate this metric at the end of every time window T. If we find that this metric is above a threshold H then we provide feedback to thread arbiter/picker to lower the priority of the thread. We keep observing this metric. If it becomes lower than the threshold H then we provide feedback to the arbiter to restore the priority of the thread. We can also introduce a 2-bit hysteresis so that we only lower priority if we observe AverageBranchMispredictStall being more than a threshold twice in consecutive time intervals and restore the priority only if we observe AverageBranchMispredictStall being less than a threshold twice in consecutive time intervals.

## 3 Methodology

I am using a qualitative methodology to compare the contributions of this paper with other existing approaches. Particularly I compared with hardware solutions and software solutions and the metrics I used are:

i) Need software support: That is, does the approach require changes from software or will it work seamlessly with existing software.

ii) Flexibility: That is, can the idea be improved or configured later on. Either software or hardware/software hybrid solutions have this advantage.

iii) Verification complexity: Hardware solutions generally have verification complexity. They need to be fully verified before they can ship. Whereas software solutions can be potentially patched.

This work can be extended by doing a quantitative evaluation with various workloads.

## 4 Results

### 4.1 Hardware solutions:

**Better branch predictors**

We can reduce mis-speculation by using better branch predictors [1]. However for integer intensive workloads like SPEC-INT there are always benchmarks which are going to be limited by branch prediction accuracy.

**Shallower pipelines**

We can reduce mis-speculation penalty by using shallower pipelines. However this will unnecessarily limit clock frequency of the processor and can negatively impact performance.

**4.2 Software solutions:**

**Compiler optimizations**

We can use compiler techniques such as loop unrolling and loop fusion to reduce the number of branches [2, 3] thereby reducing mis-speculation. These are helpful if the underlying processor does not use hardware branch prediction or uses very naive prediction.

| Solution | Need software changes | Flexibility | Hardware Verification complexity |
|---|---|---|---|
| Branch mis-speculation based feedback | No | No | Yes |
| Better Branch Predictors | No | No | Yes |
| Shallower pipelines | No | No | Yes |
| Compiler optimizations | Yes | Yes | No |

**Table 1: Comparison of Various Solutions**

**5 Related work**

I have already discussed some related work in the previous section. In this section I am going to discuss some more. There are many optimizations that have been proposed to improve performance of multithreaded processors

In [4] Snavely and Tullsen proposed co-scheduling of specific set of threads to improve overall performance. Their approach did not require any advance knowledge of the application. Their focus is on optimizing scheduler for overall performance whereas ours is a pure hardware solution which focuses on reducing mis-speculation thus leading to better resource utilization and power savings.

In [5] Tullsen et al. proposed solutions for various bottlenecks for a more realistic implementation of Simultaneous Multithreading Processor (SMT) processor. Particularly they proposed and evaluated various fetch and issue unit optimizations that will improve performance of SMT processors. Their branch counting policy (BRCOUNT), which reduces processing of wrong path instructions provide moderate speedups. For their workloads it improved throughput by as much as 8%. They mention that it is important to reduce wrong path instructions for improved performance although it is more important for a single-threaded processor than a SMT processor. However, they had only two workloads from SPEC-INT in their mix of eight program workload. The throughput benefit for BRCOUNT could potentially be more for an integer intensive workload. Further reducing wrong path instructions also improves performance-per-watt metric which is not a particular focus of their paper.

In [6] Settle et al introduced a novel way to reduce penalty introduced by inter-thread conflicts in the cache system. By using hardware Activity Vectors to monitor the access patterns of the caches and providing that information to the operating system job scheduler they reduce the inter-thread cache conflicts and improve overall performance. Their approach is focused towards improving performance by reducing cache conflicts whereas ours is a hardware solution that works seamlessly with existing software and focuses on reducing branch induced mis-speculation in the system.

In [7] Hily and Seznec showed that the performance of multithreaded processors is limited by the cache hierarchy, in particular, by contention at L2 cache levels. Their study is more focused on limitations caused by cache hierarchy on performance of SMT processors. Whereas our solution is focused on ameliorating limitations caused by branch induced mis-speculation in the system.

## 6 Conclusions

Major chip manufacturers have all introduced Multithreaded processors. These processors are used for running a variety of workloads. Efficient resource utilization is an important design aspect in such processors. Depending on the workload, mis-speculated execution can severely impact resource utilization and power utilization. In this paper I have presented a mechanism to monitor phase based behavior of branch mis-prediction stalls and use that information as a feedback to adapt the thread-arbiter pick policy. I have presented a qualitative evaluation of our approach. The solution presented in this paper does not require any software modifications. It works seamlessly with existing software.